Pacific Northwest National Laboratory
Operated by Battelle for the U.S. Department of Energy

# GridWiseTM: The Benefits of a Transformed Energy System

L. D. Kannberg
D. P. Chassin
J. G. DeSteese
S. G. Hauser
M. C. Kintner-Meyer
R. G. Pratt
L. A. Schienbein
W. M. Warwick

## DISCLAIMER

# GridWise™: The Benefits of a Transformed Energy System

L. D. Kannberg
D. P. Chassin
J. G. DeSteese
S. G. Hauser[a]
M. C. Kintner-Meyer
R. G. Pratt
L. A. Schienbein
W. M. Warwick

(a) Utility Automation Integrators

# Summary

The demand for electricity is expected to continue its historical growth trend far into the future and particularly over the 20-year projection period discussed in this report. To meet this growth with traditional approaches will require added generation, transmission, and distribution, costing up to $1.4 billion/GW ($1,400/kW in year 2000 dollars) on the utility side of the meter. The amount of capacity needed in each of these categories must supply peak demand and provide a reserve margin to protect against outages and other contingencies. The "nameplate" capacity of many power system components is typically utilized for only a few hundred hours per year. Thus, traditional approaches to maintaining the adequacy of the Nation's power generation and delivery system are characterized by lower than desirable asset utilization, particularly for assets located near the end-user.

Other issues are beginning to affect the utility industry's ability to supply future load growth. The disparity between current levels of investment in generation and transmission suggests a looming crisis that creates a strong element of urgency for finding alternative solutions. In addition, any solution needs to address the cycle of boom and bust that is typical of certain sectors of the electric industry and is likely to become more pronounced as deregulation takes hold across the Nation.

The increased availability of energy information technologies can play an important role in addressing these issues. Historically, power supply infrastructure has been created to serve load as a passive element of the system. Today, information technology is at the point of allowing larger portions of the demand-side infrastructure to function as an integrated system element that participates in control and protection functions as well as real-time economic interaction with the grid. The collective application of these information-based technologies to the U. S. power grid is becoming known as the GridWise™ vision or concept.

This report presents a preliminary scoping assessment conducted to envision the general magnitude of several selected benefits the GridWise concept could offer when applied nationally. These benefits accrue in the generation, transmission and distribution components of the power grid, as well as in the customer sector, as indicated in Figure S1. The total potential benefit of implementing these technologies over the next 20 years is conservatively estimated to have a present value (PV) of about $75 billion. When estimated on the basis of a less conservative implementation scenario, the PV of these benefits is shown to essentially double.

Other benefits enabled by GridWise technologies were identified but not fully evaluated or claimed in this study. This was done to avoid, as much as possible, accounting more than once for the benefits implicit in other advantages offered by the GridWise concept. Collectively, such benefits "left on the table" have the potential to represent additional PV in the $100 billion to $200 billion range. We leave these benefits for others to evaluate more thoroughly.

While implementation costs were not considered and the error band on the total benefit value is likely to be large (possibly ± $25 billion), the major conclusion of this exercise is that the GridWise concept has the potential for great economic value and should make a major contribution to transforming the present electric generation and delivery infrastructure into the power grid of the future.

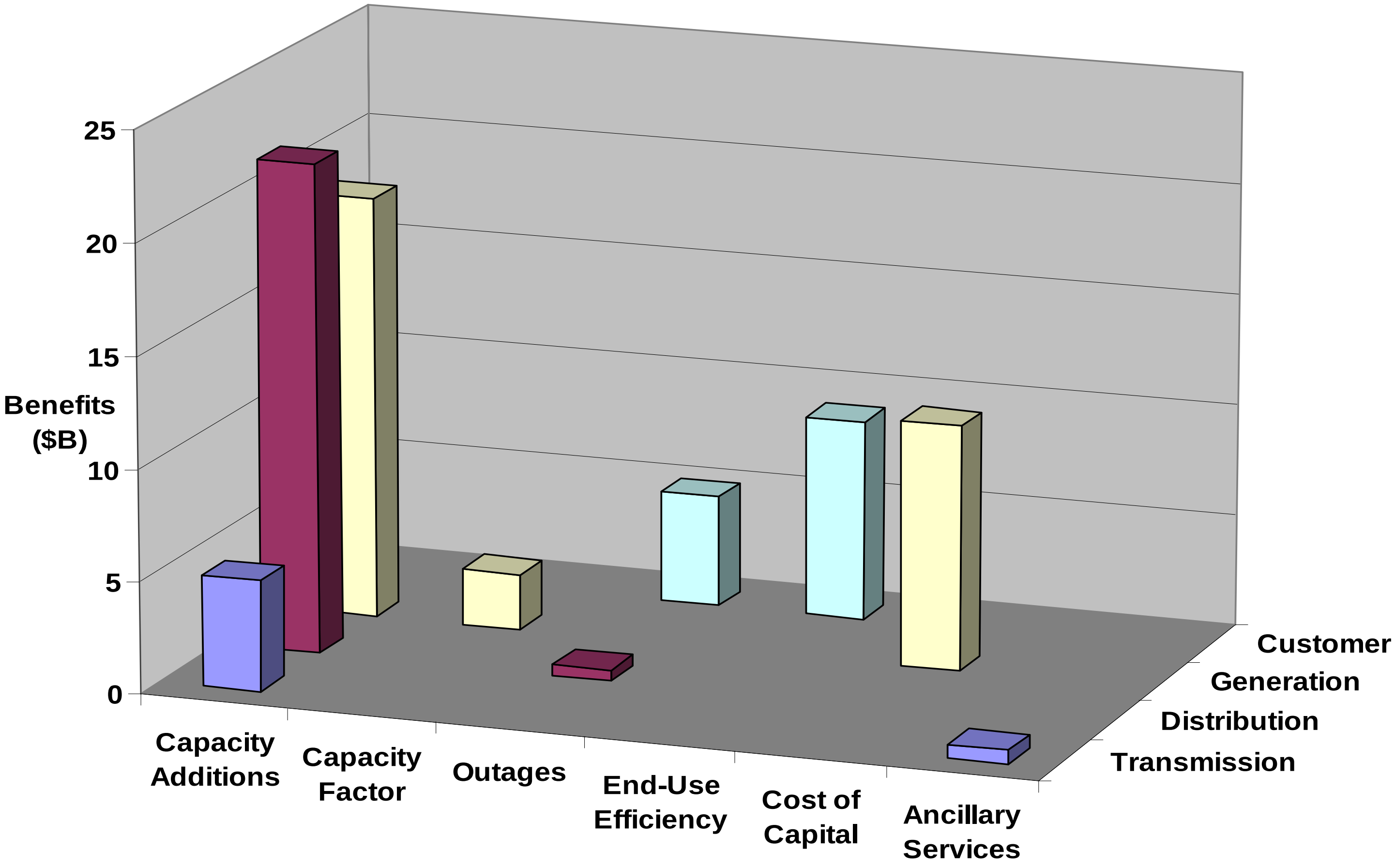

Figure S1:  Conservative estimates of the sources of GridWise benefits by utility sector

Despite the simple methodology used in this evaluation, the discovery of such large benefit values strongly encourages more rigorous analysis to determine the credible net benefit order-of-magnitude (i.e., present value of benefits minus implementation and maintenance costs) of the GridWise concept to the utility industry, end-use customers and the Nation as a whole.

# Contents

# Figures

# Tables

# 1.0 Introduction

The Nation's prosperity and the American way of life depend upon efficient and affordable energy. However, the electric power system contains many expensive and under-utilized capital assets that saddle ratepayers with a burdensome mortgage. Without a major shift in the way the energy system is planned, built and operated, the U.S. will invest hundreds of billions of dollars in conventional electric infrastructure over the next 20 years to meet expected load growth. Minimizing the cost of new electric infrastructure could be a key to strengthening the U.S. economy.

The GridWise concept for the power system of the future suggests that information technology can revolutionize electric power generation and delivery, as it has other aspects of U.S. business because, fundamentally, "bits are cheaper than iron." Bringing the electric power system into the information age would allow the Nation to realize the benefits already achieved by leading-edge industries that use real-time information, distributed e-business systems, and market efficiencies to minimize the need for inventory and infrastructure, and to maximize productivity, efficiency, and reliability.

The GridWise concept is a vision for transforming the Nation's electric system—from central generation down to customer appliances and equipment—into a collaborative network filled with information and abundant market-based opportunities. It would weave together the traditional elements of supply and demand, transmission and distribution with new "plug-and-play" technologies such as superconductors, energy storage, customer load management, and distributed generation, using information to make them function as a complex, integrated system.

With the help of information technologies and the creation of a distributed, yet integrated system, the GridWise concept would empower consumers to participate in energy markets—the key to stabilizing prices. At the same time, this transformation of the energy system responds to the urgent need to enhance national security. A distributed, network-based electric system could reduce single-point vulnerabilities. It also allows the grid to become "self-healing," by incorporating autonomic system reconfiguration in response to man-made or natural disruptions.

Implementing a GridWise infrastructure in the United States is expected to be a challenging endeavor requiring substantial resources to accomplish. Even the investment required for necessary analysis to assess the concept's potential value is large enough to justify a step-wise approach. The study documented in this report was undertaken as an initial step in this process and represents a high-level overview of the potential benefits the GridWise concept could offer if applied incrementally to the Nation's electric power system over the next 20 years. This study is based on the premise that incorporating GridWise technologies would have the primary effect of increasing utility asset utilization. This, in turn, would accrue benefits from deferral and reduced rates of new construction needed to meet anticipated load growth and from improvements in system efficiency, capitalization and energy price stability. These and other advantages of the concept including a range of customer benefits are assessed at the national level using statistics in the Annual Energy Review 2000 issued by the Energy Information Administration (EIA 2000). Benefits are estimated in year 2000 dollars unless otherwise noted.

# 2.0 Today's Electric Power System

Electric power generation, transmission and distribution are components of fundamentally a very large, just-in-time energy delivery system. At any instant, system operators attempt to control generation and the functioning of the transmission and distribution grid so that they exactly supply the total end-use load and any system losses that occur. Operating in this mode, the grid delivers power with an end-use availability of 99.97% or better in much of the United States (Willis and Scott 2000).

In 2000, the U.S. power system included over 9,000 generating stations of varying sizes with a combined net summer capacity of 819 GW. The corresponding total net generation was 3.8 x $10^{12}$ kWh. Dividing this number by the number of hours in a year (8760) indicates a year-round equivalent capacity of 433 average GW (aGW). As a measure of asset utilization, the indicated capacity factor (actual energy output over rated capacity x 8760) is about 0.53. The total book (not replacement) value of U. S. generation plants exceeded $500 billion in 1995 and was estimated at $570 billion in 2001. The time required for adding a generation asset to the system ranges from about 2 to 8 years.

The transmission system comprises over 700,000 miles of high voltage lines operating at 22 kV or greater, of which 200,000 miles are at 230 kV or greater. The capacity factor of the transmission and sub-transmission system is not precisely known but believed to about 0.5. The total book value exceeded $56 billion in 1995 and was expected to be $64 billion in 2001. The time needed for adding a major line is 5 to 10 years, as governed principally by environmental regulations and permitting procedures. However, modifications that reinforce, reconfigure or add capacity to the existing infrastructure often can be accomplished in less time.

Distribution systems total over 1 million miles of lines owned by over 3000 different utility entities (public and private). The capacity factor of this network is not precisely known, but believed to be 0.3 or less on average. The total asset value exceeded $140 billion in 1995 and was expected to approach $160 billion in 2001. The time needed to add distribution assets is typically 1 to 2 years.

The above infrastructure supplies energy to all the electrical loads on the system. Until recently, the end-use sector has generally been considered as the "passive" (i.e., demand to be served) component of the electric power system. About 125 million customers consume 3.6 x $10^{12}$ kWh annually. A typical residential customer has approximately 1/2 mile of wire installed inside his own property. The total asset value of end-use electric distribution exceeds $1 trillion and typically operates at a capacity factor of less than 0.1.

The above summary illustrates the essentially monotonic falloff in asset utilization with distance into the grid, as measured from the generator to end-user. This is a natural artifact of requiring every link and node in the grid being sized to accommodate the anticipated peak load at that location regardless of its duration. In the GridWise concept, it will be shown that the customer side of the meter becomes an active system component that creates opportunities for better asset utilization, system management and control solutions that would not otherwise be available.

## 2.1 Daily Load Curves

System load varies during the course of the day reflecting the cyclic or intermittent nature of most end-use loads. At the system level, individual loads accumulate into the typical load shape illustrated in Figure 1. In a traditional, regulated utility environment, base load plants, that are generally utilized close to the limits of their operational availability, meet the portion of the system load below the trough of the curve occurring between 4 and 5 AM. Intermediate load plants that are committed and dispatched in economic order, as needed, supply the middle tier of the load curve. Peaking plants that operate typically for only a few hours per day are also dispatched as necessary to supply the remaining portion of the peak load. Additional capacity provides a reserve margin to protect the system from contingencies such as unplanned generation and grid outages or unanticipated demands being placed on the system. In 2000, the national summer and winter capacity margin was 14.6% and 26.9%, respectively. Thus, the total asset capacity available must be sufficient to supply the highest load of the season plus the required reserve margin for that load.

The daily system load curve is an aggregation of a large number of daily distribution load curves, which are, in turn, aggregations of individual end-use load shapes such as the representative commercial building demand curve shown in Figure 2.

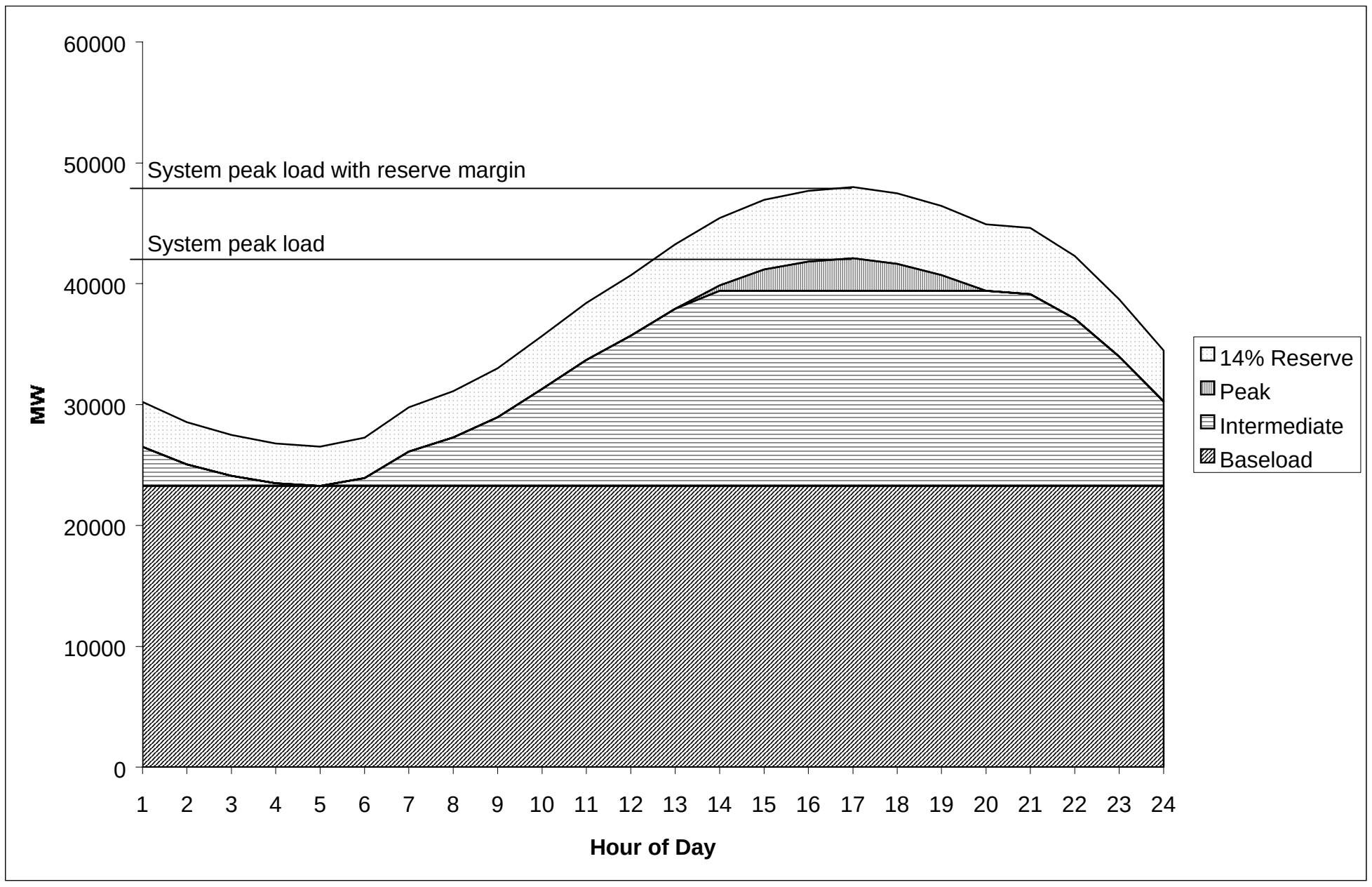

Figure 1**:** ERCOT (Texas) daily system load curve on August 27, 1990 illustrates a typical commitment of generation assets. The baseload is level all day at roughly 23.3 GW. The intermediate load goes up to approximately 16.1 GW and peak load increases another 2.7 GW. The reserve capacity ranges from 3.3 GW up to roughly 5.9 GW.

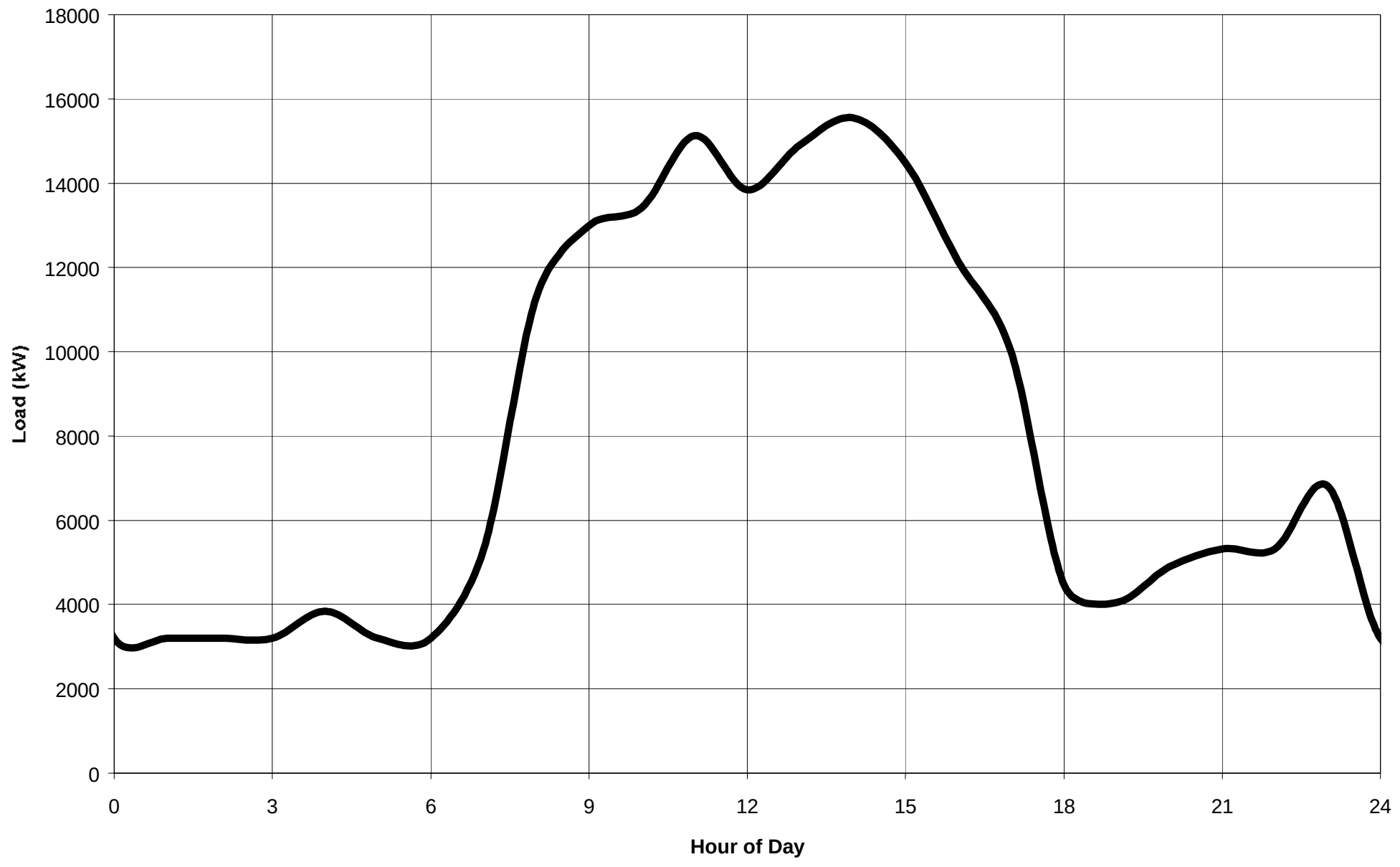

Figure 2**:** A typical Pacific Northwest commercial building[(1)] has a load curve shape quite different from that of the system.

This example illustrates that the load shapes of individual end-users are less predictable than the system load shape. However, at the system level, the natural diversity of individual demand profiles is integrated to form the much more predictable and, therefore, manageable load shape shown in Figure 1.

## 2.2 Annual Load Duration Curve

A load duration curve (LDC) is a valuable means of displaying the asset utilization of utility system components. The annual LDC is a histogram of the system load factor (LF) over an entire year, sorted in order of descending LF. Figure 3 shows 1993 LDCs at the system generation level and for a typical distribution substation of the Pacific Gas and Electric Company (PG&E). These curves represent LF as a percentage of rated capacity versus percentage of the year (8760 hours) and illustrate how much of nominal capacity is utilized throughout the year. The figure shows that there are only a few hours in the year when the rated capacity of these assets is fully utilized. As discussed previously, the distribution system is not as well utilized as generation assets.

[(1)] End Use Load and Consumer Assessment Program (ELCAP) Site 444, a bank building in Seattle, on Wednesday, July 29, 1987.

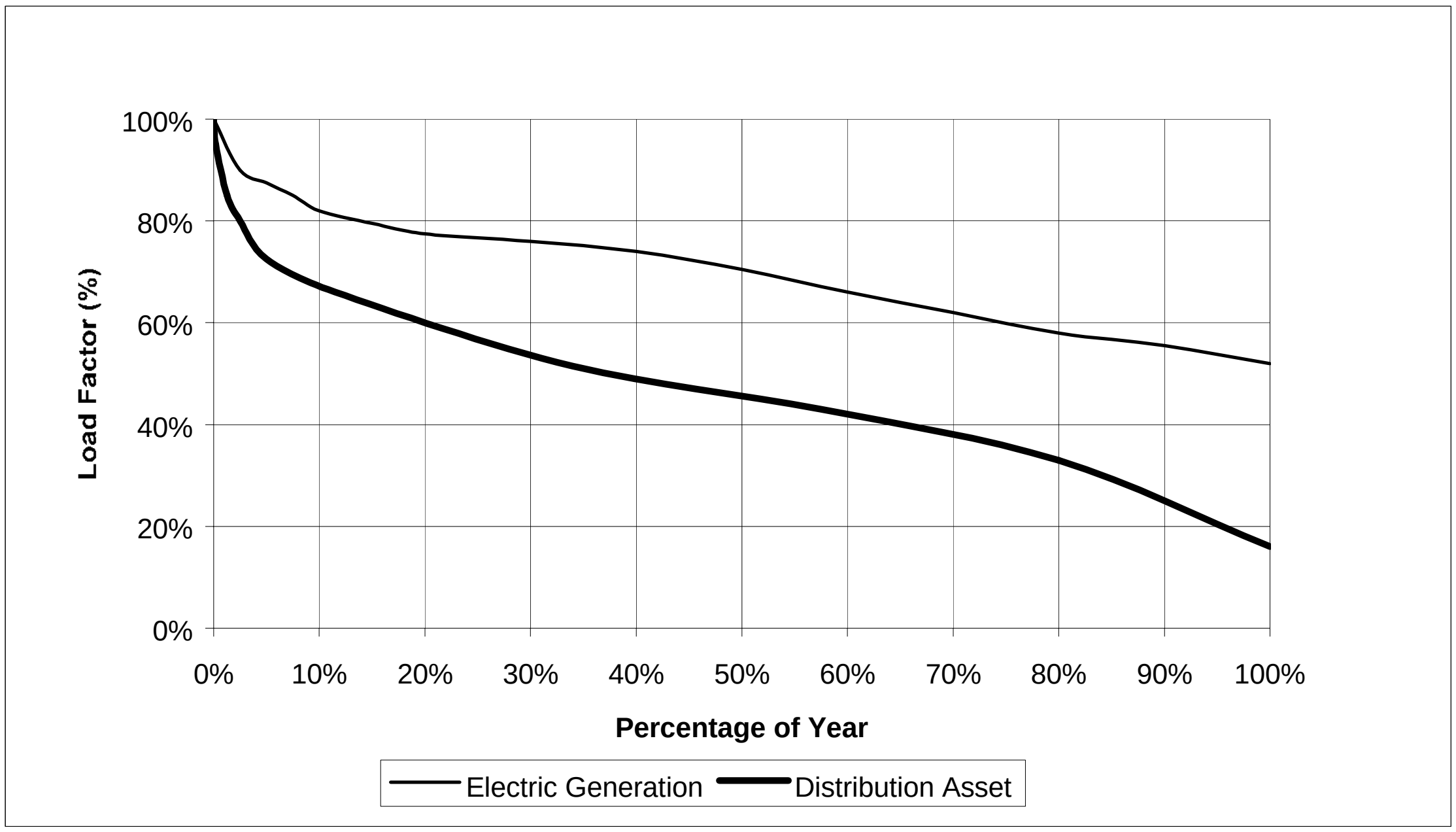

Figure 3: PG&E load duration curves for 1993 illustrate decreased utilization of assets closer to the customer (EPRI 1993).

## 2.3 Generation

Generation assets in a variety of types including coal-, oil-, gas- and nuclear-fueled steam turbines, oil- and gas-fueled combustion turbines, combined cycle units and hydropower provide most of the Nation's electric power. Other sources utilized are power imports and a small but growing contribution from renewable sources such as solar, wind and biomass energy. With the exception of a few battery stations, the grid has no significant means of storing energy electrically. Historically, the majority of electric utilities were vertically integrated businesses in which individual companies owned and operated all aspects of power generation and delivery. In this mode of ownership, generation resources were operated primarily to serve the reliability requirements associated with the utilities' obligation to serve loads on the system. However, in recent years, industry deregulation (and re-regulation) has started the dismantling of the traditional utility business structure so that generation, transmission and distribution are becoming owned and operated by different entities. As a result, electricity is now sold as a market commodity and the focus of generation operations has shifted much more toward achieving the maximum economic efficiency.

There has been a steady decline in generation capacity margins since 1980 (see Figure 4). In 2000, the U. S. power system operated with a summer reserve margin of 14.6%. Figure 5 shows forecasts made by the EIA and the North American Electric Reliability Council (NERC 2002) that show a sharp increase in planned capacity additions in the years immediately following the 1999 - 2000 power crisis. However, of the 43-GW capacity growth planned between 1995 and 1999, only 18 GW were built. The capital cost of added generation assets averages $600/kW or about $9 billion per year at EIA-projected growth rates.

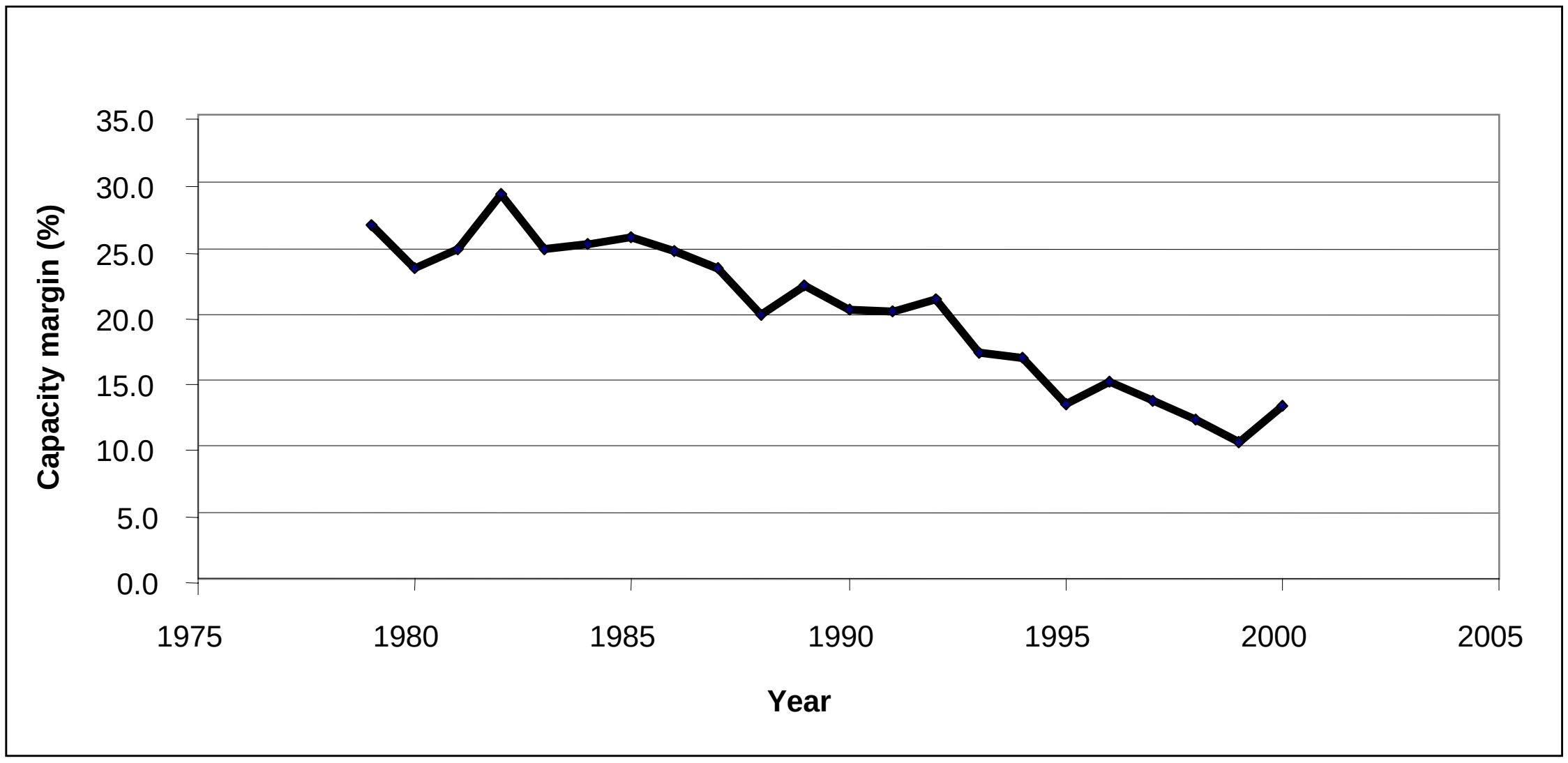

Figure 4: Summer capacity margin from 1979 to 2000 shows a steady decline over the years before the power crisis of 1999 and 2000, when it dipped below the present 14.6% needed for reliable operations.

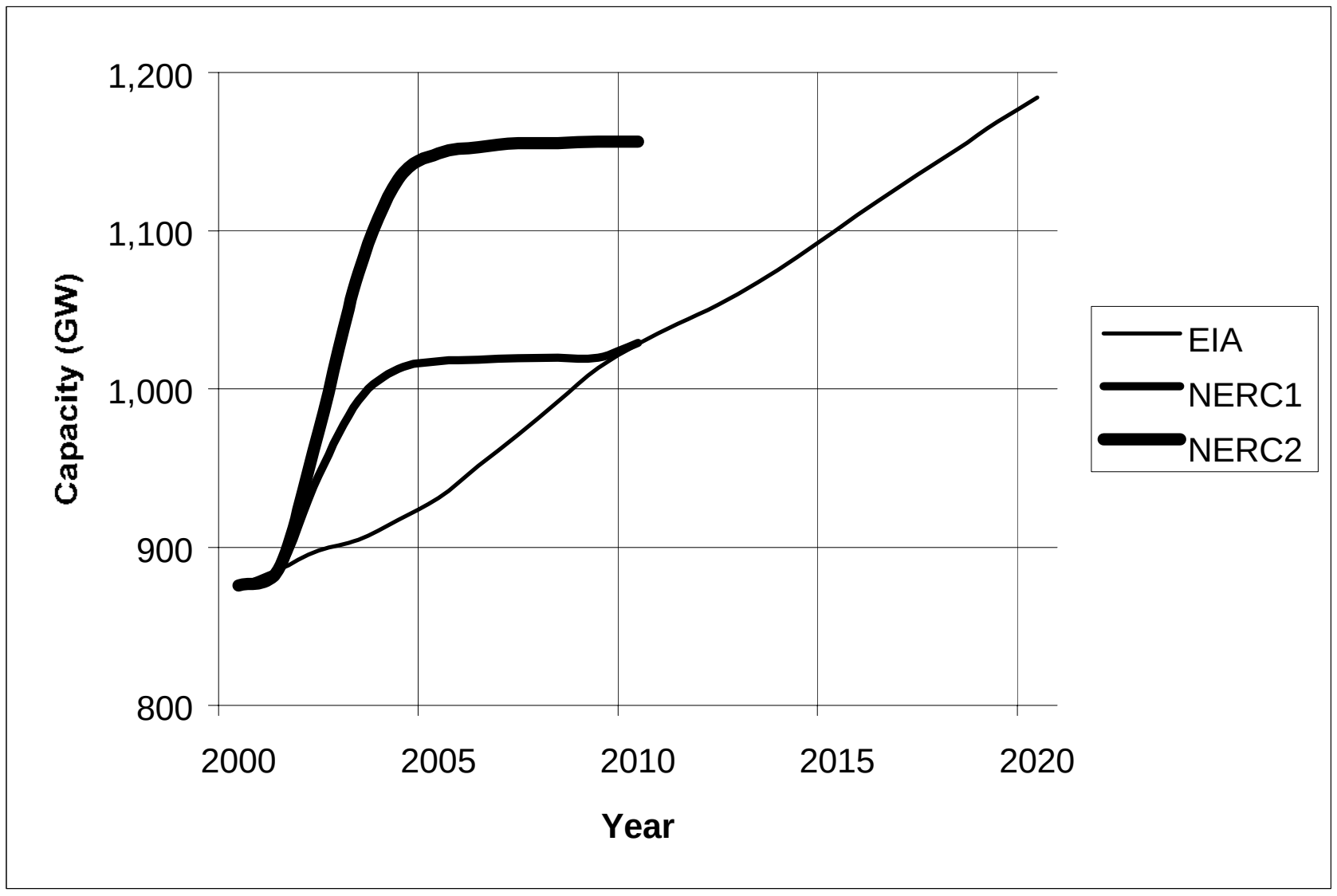

Figure 5: EIA and NERC forecasts of North American generating capacity growth from 2002 to 2020. The NERC forecasts are based on permitting and construction announcements that are not projected beyond 2010. The forecast NERC 1 represents existing capacity plus the best estimate of future additions by Energy Ventures Analysis, Inc., acting as an advisor to NERC's Reliability Assessment Subcommittee. Forecast NERC 2 is a projection of existing capacity plus all announced merchant generation additions from information supplied to NERC by the Electric Power Supply Association.

## 2.4 Transmission

Power from generating plants is delivered to transmission substations, where it is transformed to high-voltage electricity for transmission over long distances. Typical transmission voltages in the U. S. include the extra-high voltage of 765 kV, and the 500 kV, 345 kV and 230 kV voltages of the most common long-distance transmission lines. In a few instances, bulk power is also transmitted as high-voltage DC requiring AC to DC, and the reverse DC to AC conversions, respectively, at line terminals so that the DC power can be received from and delivered to the AC grid. Other common transmission voltages include 161 kV, 138 kV, 115 kV and 69 kV. Of these, the lower voltages, 115 kV and 69 kV, are sometimes called sub-transmission voltages. Sub-transmission refers to a lower level in the grid hierarchy that typically interfaces the long-distance bulk transmission backbone with the distribution network supplying end-use customers.

In the decade 1979 to 1989, transmission capacity grew at rates commensurate with the growth of generating capacity and summer peak demand. Since then, transmission system expansion has lagged behind demand growth rate and is expected to continue to do so into the future. The percentage growth rates of electric power transmission and summer peak demand that occurred in the U. S. between 1988 and 1998 and those predicted for the period 1999 to 2009 are illustrated in Figure 6 (Hirst and Kirby 2001).

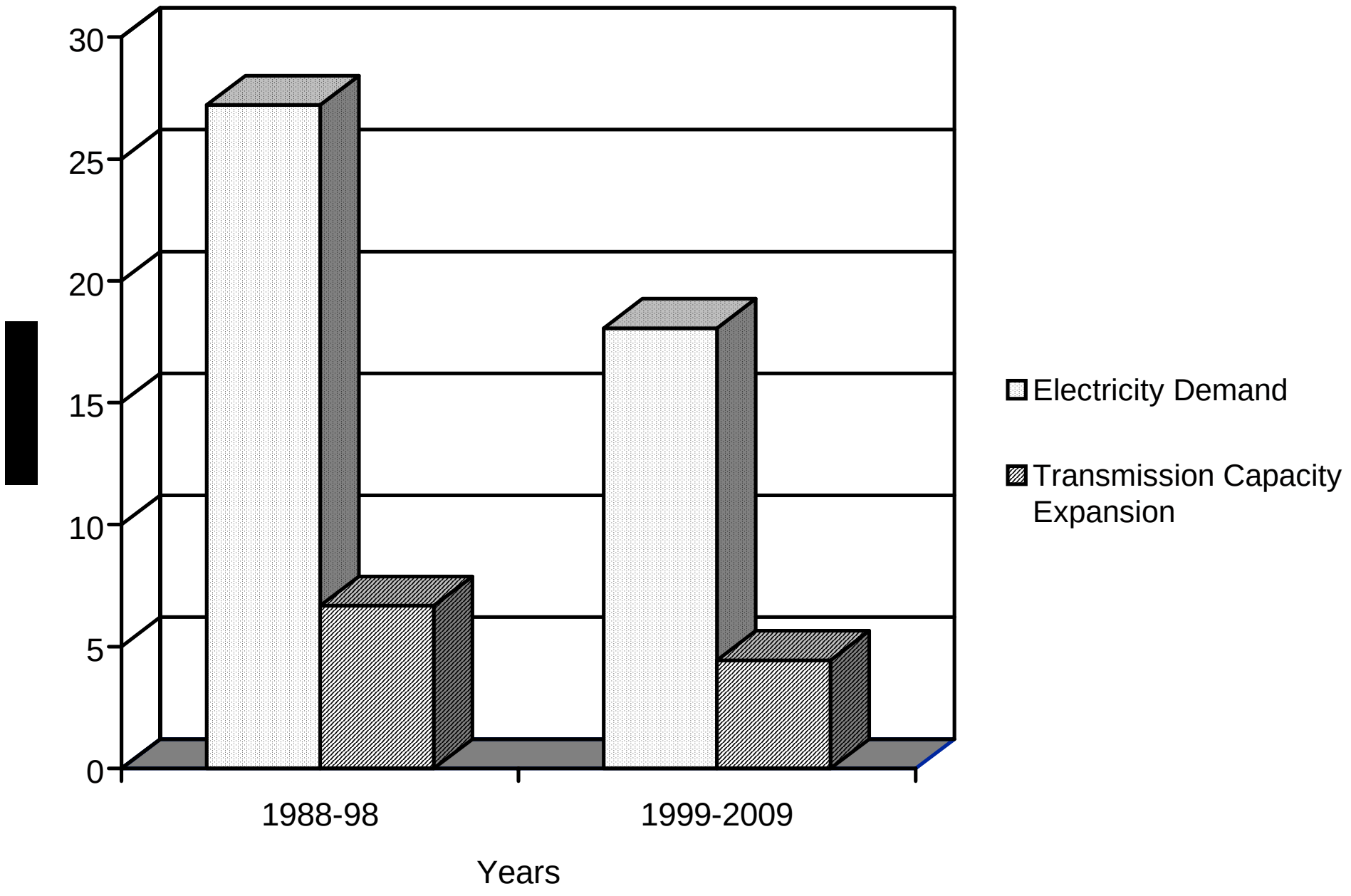

Figure 6: Transmission capacity expansion lagged behind generation growth rate in the last decade and is projected to continue this trend.

A similar comparison is made in Figure 7, where the planned transmission growth forecast for North America is much lower than transmission growth rates that correspond to the EIA and NERC forecasts of generating capacity growth shown in Figure 5. Approximately 10,400 GW-miles of new transmission assets are planned between 2002 and 2010, but as much as 26,600 GW-miles (Hirst and Kirby 2001)[2] may be needed at a cost of approximated $28 billion. Transmission additions would cost roughly $3.9 billion[3] per year if installed to maintain parity with projected generation growth rates. In contrast, only $115 million was invested in transmission capacity expansion in 1999.

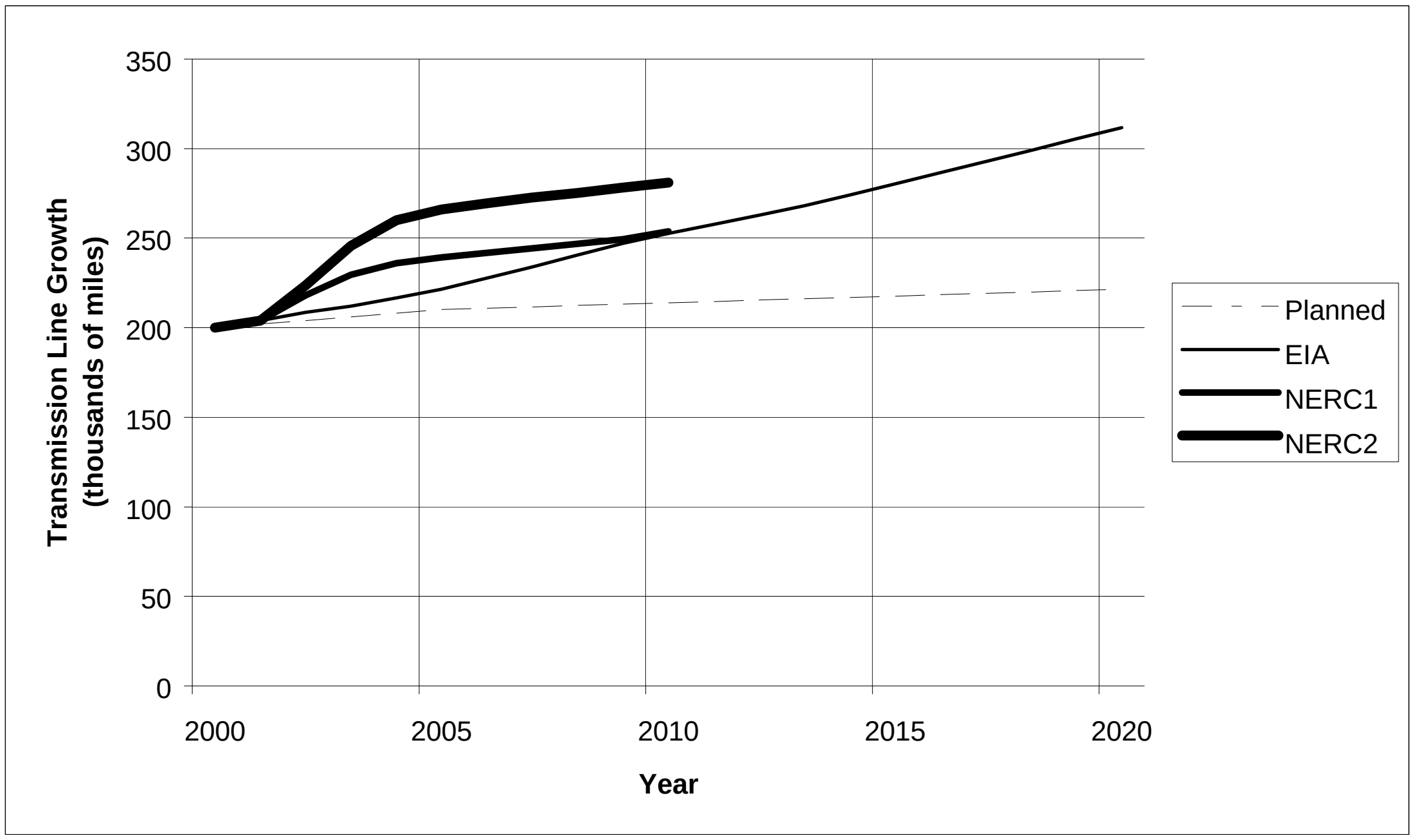

Figure 7: Transmission growth forecast for North America is much lower than growth commensurate with generation growth projected in Figure 5.

## 2.5 Distribution

The distribution system is the infrastructure that delivers power to end-users from substations supplied by the transmission system. At these substations, power delivery is reduced from the high-voltage levels suitable for long-distance transmission to lower voltages appropriate for local distribution. Typical distribution systems operate at 34.5 kV, 14.4 kV, 13.8 kV, 13.2 kV, 12.5 kV, and sometimes lower voltages. With only a few exceptions, distribution systems are designed using a unidirectional radial power-flow topology composed of feeders and laterals. Radial feeder lines typically fan out from each substation and, in turn, supply power to lateral lines that feed end-use service transformers. Except in very high population density areas, there is generally little redundant network structure at the feeder level because of the prohibitive cost (i.e., there is seldom a redundant supply path to most end-use loads).

---

(2)Hirst and Kirby considered a normalized ratio of 201 MW-miles/MW of demand to arrive at this value.

(3)Assuming average transmission cost of $1 million/GW-mile and the need for 150 miles of transmission infrastructure for every GW of added generation capacity.

The distribution system is designed to maintain feeder voltages in conformity with the standard ranges defined by the American National Standards Institute (ANSI 1995) when supplying the maximum connected load. Thus, distribution infrastructure, including substations; feeders and laterals; service transformers; connections and meters, must be continuously expanded to keep up with end-use load growth at a cost of approximately $685 million/GW. Service transformers, connections and meters constitute approximately 50% of total distribution system cost. At currently projected growth rates, the investment required to maintain adequate distribution infrastructure could be as high as $9.4 billion per year for the next 20 years. Distribution system load factors are typically less than 40%.

## 2.6 Customer

Service transformers reduce distribution voltage to the final customer supply level. The secondary windings of service transformers connect directly to customer facilities, completing the power flow path from generating plant to the customer. The electric power grid typically operates as a three-phase network down to the level of the service transformer. Some industrial and commercial customers are supplied with three-phase power, while residences are generally supplied by single-phase connections. Service transformers feeding single-phase loads are connected in a manner that is designed to balance the total load on each phase of the three-phase distribution system.

In 2000, electric retail sales amounted to $3.4 \times 10^{12}$ kWh and produced revenues of approximately $215 billion. End-use electricity consumption was split in roughly equal proportions between industrial ($1.07 \times 10^{12}$ kWh), commercial ($1.03 \times 10^{12}$ kWh) and residential ($1.19 \times 10^{12}$ kWh) customers (EIA 2000).

## 2.7 The Cost of Load Growth

The total capital cost of infrastructure needed to supply an additional 1 GW of load ranges from $600 million to $1.4 billion on the utility side of the meter alone. This range comprises the sum of $600 million/GW for generation, up to $150 million/GW for transmission, and up to $685 million/GW for distribution. Thus, an annual capital investment in the range between $8.2 billion and $19.2 billion will be required to accommodate the presently forecast U. S. demand growth rate of 13.7 GW/year.

The capital cost of connecting electrical appliances and equipment is also substantial on the customer side of the meter. For example, industrial customers are estimated to invest approximately five times the utility’s distribution cost per unit of capacity to distribute power and protect equipment inside their facilities.

# 3.0 Evolution Toward the Power System of Tomorrow

The transformed energy system that would be enabled by implementing the GridWise vision has the potential for fundamentally altering the way the Nation's electric power system functions. As reviewed later in this report, the resulting annual benefits should grow at rates exceeding $5 billion with a present value of about $75 billion or more over 20 years. This section identifies the principal benefits and attributes of the concept that are discussed in more detail in the following sections.

## 3.1 Basic GridWise Benefits

There are four main benefits expected from a transformed energy system:

1. Existing assets can better perform their current functions, e.g., generating plants meet load more efficiently.

2. Existing assets can perform new functions, e.g., backup and on-site generation serve feeder loads or provide services such as transmission reliability functions.

3. Existing assets can be deployed to provide existing functions, e.g., load provides ancillary services.

4. New assets can perform new functions, e.g., load function arbitrage and balancing can be performed at the customer or feeder level.

The above benefits would derive from a variety of GridWise attributes and values. These include:

- higher asset utilization permitting system operators to provide more services with the same installed capacity and install less new equipment to meet the same growth

- increased efficiency provided primarily by a flatter load duration curve, increased investment opportunities in end-use efficiency improvements, and increased use of combined heating and power (CHP) systems

- improved system operations through more effective sources of ancillary services (see Appendix Section A.3), improved energy security and higher quality power

- avoided costs realized through lower cost of capital resulting from lower risk economics, reduced maintenance costs and shorter outages

- energy price stability and predictability achieved by increase demand elasticity

- intangible social benefits such as decreased customer discontent, greater personal and economic security, and greater confidence in public governance.

## 3.2 Just-the-Right-Size System Capacity

A principal benefit of the GridWise concept would be to enable significant improvements in power system asset utilization. Consequent benefits to generation include deferrable capacity additions realized by reducing necessary reserve margins and increasing existing capacity factors. Figure 8 illustrates these values as applied to California's load duration curve of 2000.

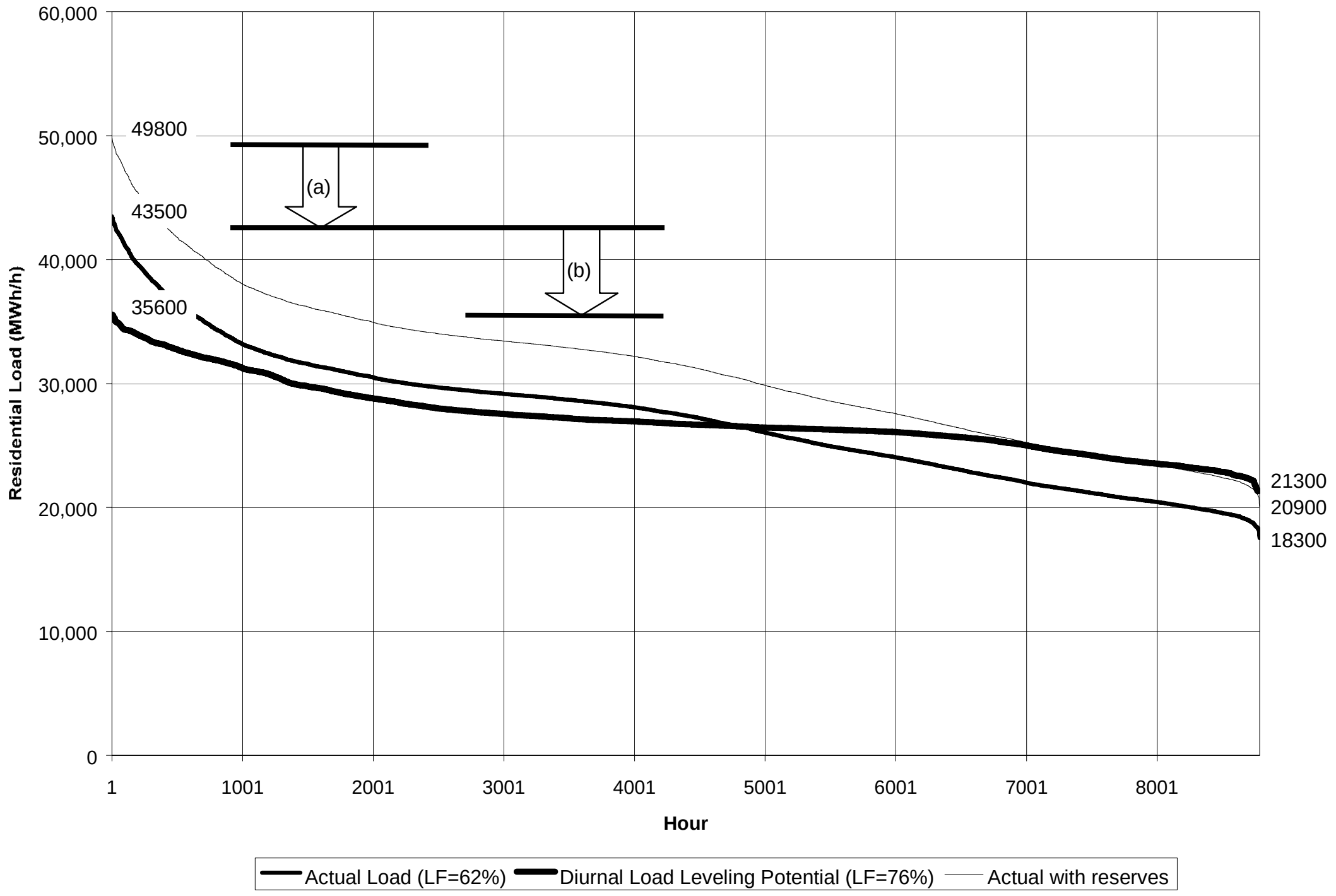

Figure 8: Potential values of reshaping California's Year 2000 load duration curve. Principal benefits are designated: (a) reduced need for reserve capacity and (b) increased asset utilization.

Benefit (a) is obtained by using load to provide ancillary services that reduce the amount of reserve margin needed. Benefit (b) is obtained by shifting load to off-peak hours of the day. In Figure 8, the system load factor without load management is 62%. With load management, the load factor could rise to as much as 76%, with a peak capacity reduction of about 22%. Scaled up to the national level, this would result in a reduced need for capacity additions of about 135 GW, worth approximately $80 billion in capital savings.

With respect to long-distance power transmission, GridWise technologies are expected to enable reductions and deferrals of new capacity commensurate with reduced rates of generation capacity addition indicated above. In addition, the increased use of “grid-friendly” load as an active control measure can increase the reliability and security of the power delivery system.

Distribution systems realize benefits by virtue of the substantial costs required to expand capacity. There is a marked tendency to overbuild systems when expansion does occur, resulting in notably low capacity factors. With added flexibility provided by GridWise enabled loads, significant uncertainty in load growth can be reduced or eliminated.

The ability of GridWise technologies to offer a fundamentally new approach to grid asset utilization and the benefits of widely distributed “active” as opposed to “passive” end-use load can result in a significant future transformation of the Nation’s electricity generation and delivery system. The following sections describe the order-of-magnitude assessments we made to evaluate these benefits.

# 4.0 Power Generation Benefits

Power generation would benefit from implementing GridWise technologies in several ways including increased revenues, lower investment and operating expenses, and lower capital risk and uncertainty.

Increased revenues would result from increased commodity sales, increased access to markets (from reduced congestion and growing numbers of local markets), increased opportunities for bilateral agreements, and increased productive generating plant life.

Lower expenses result from decreased or deferred capacity investments, lower and more stable fuel costs and emissions, and more optimal plant design and operation. Operating a plant at higher capacity factors also reduces the plant's revenue requirements. Among these, the ability to defer capacity investment and the capacity factor revenue benefit were evaluated in this study.

Lower risk and uncertainty results from more stable prices and capital markets, fewer unscheduled outages and costs, more long-term contracts, and decreased fuel price volatility. Among these, only the avoided cost of capital was evaluated.

## 4.1 Generation Deferral Benefits

The estimation of generation deferral benefits is based on two scenarios described in detail in the Appendix, section A1. In principle, both scenarios are based on the expectation that the gradual penetration of GridWise technologies will tend to flatten the shape of the national load duration curve. Scenario 1 is the more conservative of the two and assumes that 111 GW of currently excess generating capacity could be released to supply load growth and offset unit retirements in the ideal, but impractical, case of achieving a perfectly flat load duration curve. The 20-year aggregate penetration of GridWise technologies is assumed to be 50%. This is equivalent to the more realistic expectation that a completely flat load duration curve is not achievable in practice. Scenario 1 maintains the present summer reserve margin of 14.6% but assumes average generation capacity factor increases to 70%. Technology penetration growing to 50% in 20 years would defer construction of 2.8 GW of generation capacity annually (i.e., 0.5 x 111 GW released over 20 years). The benefit in avoided capital investment is $1.7 billion annually with PV of $19 billion over 20 years. It should be noted that trade-offs can be made between capacity factor, reserve margin and technology penetration rate. Any other combination of assumed values for these variables that results in an annual deferral of 2.8 GW for 20 years is an equivalent scenario.

In Scenario 2, the same technique was applied to evaluating the capacity deferral benefit based on a demand release of 285 GW, a reduction in the reserve margin to 5% and the average capacity factor increasing to 85%. The justification for this more optimistic scenario is detailed in the Appendix, Section A.1. Again, assuming a $600/kW cost of new generation capacity and a 50% GridWise technology penetration over 20 years, 7.1 GW of capacity (0.5 x 285 GW over 20 years) worth $4.3 billion is deferred annually. The present value of this benefit is $49 billion.

## 4.2 Avoided Capital Risk Benefit, 2010-2020

The estimation of avoided capital risk and accrued savings was based on EIA projections for capacity additions from 2010 to 2020 (EIA 2000). The risk reduction was evaluated by changing the average bond rate from 10% to 9%. We assumed that this benefit would not start accruing before a substantial penetration of GridWise technologies had already occurred. This is because investors would need to appreciate that the concept actually reduces financial risk before reflecting this knowledge in lower interest rates. To accommodate this projected situation, we estimated the benefit only for the 10-year period between 2010 and 2020.

For each year in this period, the capital cost of new generation was estimated reflecting the future additions mix projected in EIA 2000. The cost of new generation capacity was assumed to be $600/kW for fossil plants, $300/kW for combustion turbine plants, $500/kW for combined cycle plants, and $1000/kW for all other types of plants. The cumulative value of issued bonds amounted to approximately $80 billion at the end of the 10-year period. The PV of a 1% reduction in bond interest rate over this period was estimated at $11 billion in year 2000 dollars.

The GridWise concept can be expected to provide system operators with a greatly enhanced ability to dispatch and manage large numbers of smaller power plants including generation connected into distribution systems (distributed generation - DG) and thereby located nearer to the end-use loads they serve. Small DG additions can be constructed to track smaller increments of load growth. Shorter deployment lead times are a key economic benefit of smaller capacity additions. The ability to adjust the construction schedule in response to market changes yields improved financial performance, in excess of 4.5 times better, with only a 2.5-year lead time over the referenced 15-year base case (Lovins et al. 2002). Smaller DG units often have lead times much shorter than 2.5 years and can be expected to be even more financially attractive. These values reduce the risk investors take in financing new construction. As a direct result, the cost of capital will be lower.

## 4.3 Generation Capacity Factor

A higher capacity factor improves the profitability a generating plant. Excess plant revenue is calculated on the basis of the return requirement on capital. This is about 8.7% ($52/kW when the plant capital cost is $600/kW) plus a fixed operations and maintenance (O&M) cost of about $27/kW-yr. In Scenario 1, the benefit of deferring 2.8 GW annually is worth $221 million ($79/kW-yr x 2.8 x$10^6$ kW) with 20-year PV of $2.5 billion. Deferring 7.1 GW annually in Scenario 2 has a value of $561 million with 20-year PV of $6.4 billion.

# 5.0 Power Delivery System Benefits

Connecting generation with end-use load, the transmission and distribution (T&D) system constitutes the power delivery component of the grid. GridWise technologies offer a number of benefits to the T&D system. These include increased sales resulting from increased capacity factor, increased market transactions (from reduced congestion and growing numbers of local markets), increased asset utilization and productive asset life, fewer outages and greater revenue growth. Among these benefits, only the reduced outage benefit was evaluated in this study.

Another group of T&D benefits include lower expenses, lower capital risk and reduced uncertainty. Lower expenses result from decreased or deferred capacity investments, reduced ancillary services cost, and more optimal system planning, design, and operation. Additional benefits include decreased losses and fewer hours of over-capacity operations. Among these, only capacity deferral and ancillary service benefits were evaluated.

Lower risk and uncertainty are associated with more stable prices and capital markets, fewer unscheduled outages, increased inherent system stability, and fewer stranded assets stemming from more reliable load growth projections. None of these were evaluated in this study.

## 5.1 T&D Outage Reduction Benefits

Of outages experienced in a typical year, approximately 20% result from transmission system defects and 80% are caused by distribution problems. Assuming a 50% penetration of GridWise technologies would eventually enable a 50% reduction in transmission outage frequency, the corresponding avoided lost revenue is worth $3 million annually, with a PV of $34 million over 20 years. The remaining 80% of outages result from distribution failures. With the same reduction in outage frequency and technology penetration, the avoided revenue loss is $48 million annually, with a 20-year PV of $550 million. It should be noted the corresponding customer benefit is approximately an order of magnitude larger (see Section 6.5).

## 5.2 T&D Capacity Deferral Benefits

The T&D capacity deferral benefit was assessed on the same two-scenario basis as used to estimate corresponding generation benefits. This is justified by the general expectation that T&D infrastructure must grow at some rate proportional to the growth in generation capacity, as discussed in Sections 2.4 and 2.5, and the Appendix, section A2.

The cost of transmission was assumed to be $150 per GW of generation additions or load growth. In Scenario 1, with 111 GW of generation deferred over 20 years at a 50% GridWise technology penetration, annual deferral of transmission additions is valued at $420 million, with a PV of $5 billion. The corresponding values in Scenario 2 (285 GW deferred over 20 years and 50% technology penetration) for deferral of transmission additions are slightly more than $1 billion annually and a PV of $12 billion.

Distribution system deferral benefits were derived similarly using the value of $685/GW for infrastructure construction cost. In Scenario 1, distribution system deferral is worth $1.9 billion annually with a PV of $22 billion. The corresponding values in Scenario 2 are $4.9 billion annually and a PV of $56 billion.

## 5.3 Distribution Benefit—Alternate Calculation

An alternate method of calculating distribution benefits was evaluated. We assumed a 5% distribution system upgrade annually (equivalent to a complete system upgrade every 20 years) and a 2% load growth rate (13.7 GW annually). A 50% penetration of GridWise technologies would result in an annual deferral 6.85 GW of distribution upgrades worth $4.7 billion when implemented. If the top 20% of the load (with an average duration of about 175 hours per year) were excluded, the resulting 8-year upgrade deferral would have a PV of $29 billion. This suggests that estimates based on Scenario 1 are conservative.

## 5.4 Ancillary Services Benefit

The grid-friendly load component of the GridWise vision can be expected to supply ancillary services that contribute to grid control and operation as described in the Appendix, Section A.3. Generation resources conventionally provide ancillary services. However, by virtue of this benefit being generated by intelligent/active load components in the grid of the future vision, this benefit was counted as a T&D benefit in the present study. In the hypothetical limit, all GridWise connected loads could provide ancillary services to the extent that they can be dispatched by the system operator. The benefit was estimated by taking the previous generation deferrals and considering the equivalent load provides 50% of the ancillary services total value for all hours of the year.

Hirst and Kirby (1996) indicate an average ancillary service cost of 0.414 ¢/kWh in 1993 dollars. Using this value as shown in the Appendix, Section A.3, the first year benefit in Scenario 1 is $50.8 million with a 20-year PV of $0.6 billion. The corresponding values in Scenario 2 are $128.8 million and $1.5 billion, respectively.

# 6.0 Customer and Other Benefits

The GridWise vision converts end-use load from a passive element into an interactive system control and feedback mechanism. This leads to a variety of customer benefits derived from a combination of increased revenues, decreased costs, and reduced risks from uncertainty.

Increased revenues are primarily from the sale of surplus on-site generation and the provision of ancillary services. Decreased expenses are also realized in the form of lower demand charges and total electric costs, reduced outage costs, increased opportunities for fuel arbitrage/switching and power generation, reduced capital equipment and interconnect costs, and reduced O&M costs. Finally reduced risk from rate and cost control, increased reliability, and additional/equitable market influence are also realized.

In addition to other benefits, GridWise technologies offer the customer an opportunity to create a price-demand response in real time and thereby exercise active control on the cost of electricity service. The customer benefits evaluated in this study are described below.

## 6.1 Price-Demand Response

The bulk supply of electricity is generally limited by the availability of generation and transmission capacity. In open markets, this limit expresses itself as an escalating price as the quantity of power demanded approaches it, as illustrated in Figure 9. When customer demand is inelastic, price fluctuations have little impact on the amount of power used. In contrast, increasing demand elasticity reduces power usage as price increases. The GridWise concept is expected to increase demand elasticity by enabling consumers to adjust their power consumption when supply is constrained and prices are high. In Figure 9, the benefit of demand elasticity (1) is not significant when supply capacity far exceeds the peak load on the system (a). However, as load grows (b) and the price of power increases, the benefit of demand elasticity (2) is greatly enhanced. Not only does the demand elasticity moderate high prices, but it also provides a much more advantageous margin of safety to operations by providing larger drops in peak load when supply is constrained than when supply is relatively unconstrained.

When the economy grows, supply becomes constrained, prices rise sharply and investors are attracted to the plant construction market. The construction process has intrinsic delays that limit economic growth and often lead to over-construction of generation. Supply prices become depressed when these plants come online, causing investment capital to flee the plant construction market. Depressed energy prices persist until economic growth catches up and constrained supply returns, restarting the cycle. What is needed is a price moderating mechanism that mitigates this boom and bust cycle. Price responsive demand is an important element to controlling such adverse cycles in plant construction and power prices.

The specific strategies employed to bring about demand response are not important. Rather their characteristic of providing a counter to scarcity rents (Stoft 2002)[4] is the key feature sought. As the economy grows, those loads that are least productive will begin to reduce their consumption in favor of those that are most productive and can afford higher energy prices. This leads to increased economic efficiency and productivity, while moderating high energy prices during the period of plant construction. When the plants eventually come online, the less productive loads may once again benefit from lower energy prices. Concurrently, price will not fall so low as to cause reduced revenues to investors in the plants.

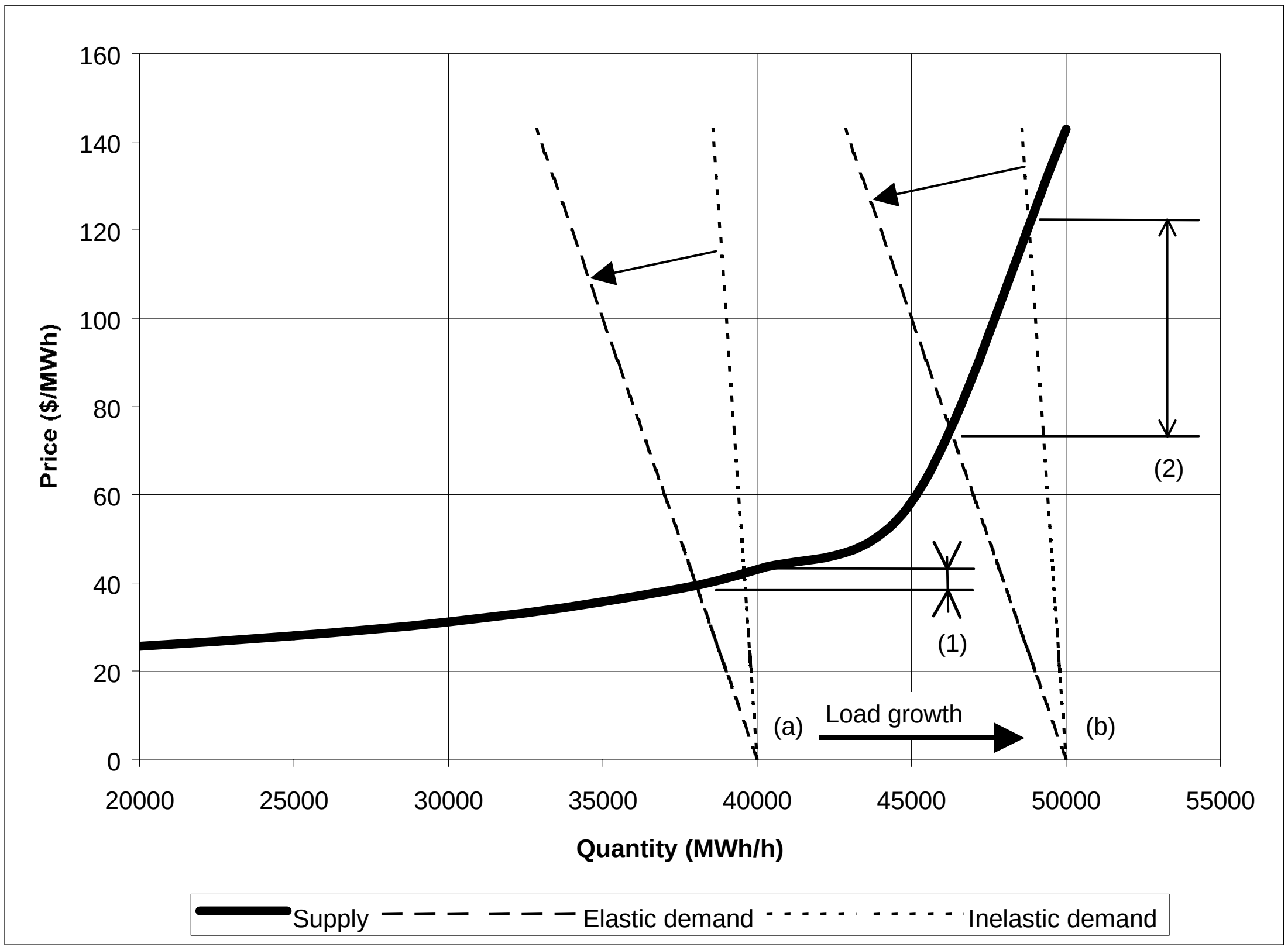

Figure 9: The ability of demand elasticity to control high prices is greatly enhanced when supply becomes highly constrained.

## 6.2 Price Volatility Impact

The impact of price volatility on customer bills was evaluated. Price volatility is characterized by very large changes in price for very small changes in supply. The impact of a 7.5-MW reduction in demand per dollar increase in price was evaluated on the basis of California Independent System Operator (ISO) operations in 2000 (Braithwait and Faruqui 2001). We estimated there would be a reduction in the peak price of $256 per MWh (42%) with a corresponding reduction in

[4] Scarcity rent is defined by Stoft as "revenue minus variable operating costs (which do not include startup costs and no-load costs)"

peak load of 2.1 GW (4.6%). This would result in a net savings to customers of nearly $1.2 billion (5.7%). The estimate for the national savings is approximately 3% reduction in electric utility costs to customers.

On a national level, we estimated that 6.0 MW/$ in demand response would result in annual reduction of nearly $6.9 billion in customer electric power bill reductions and 9.68 GW in reduced peak load. This represents a present value of $79.1 billion over 20 years.

An alternative and more aggressive approach to evaluating price volatility impact is described in the Appendix Section A.6. This approach, extrapolated to only a 1000-GW system, indicates an annual benefit of $14 million and a 20-year PV of $160 million. Because of the wide range of these estimates and the fact that they possibly may embed, at least in part, other benefits already counted, we have not included the price volatility benefit in the conservative benefit summary.

## 6.3 Enhanced Reliability and Security

Power outage-costs approaching $100 billion annually in lost economic activity have been claimed by various sources, especially by representatives of the digital economy. This value far exceeds the actual cost of the power lost. Uninterruptible power supply sales are approximately $2.6 billion annually as one indication of the value customers place on supply continuity.

Increasing the number of smaller generating units has an overall reliability benefit that can be relatively easy to quantify, even if smaller units are less reliable than the larger ones. Consider the simple case where 100 large 1000 MW generating units provide power to a region. If each unit has a costly 1% chance of failure at any given hour, then we should plan to unexpectedly lose 1000 MW of the supply about 88 hours per year, and 2000 MW for about 1 hour/year. However, if we had 10,000 units providing each 10 MW with a much more cost-effective 10% failure rate, then we can plan to lose only 20 MW of supply for 88 hours each year, and the probability of losing 1000 MW any given hour is $0.1^{100}$ or $10^{-100}$, which is essentially never. All other costs being equal, having many smaller DG units in operation leads to higher overall system reliability and lower operations and maintenance costs.

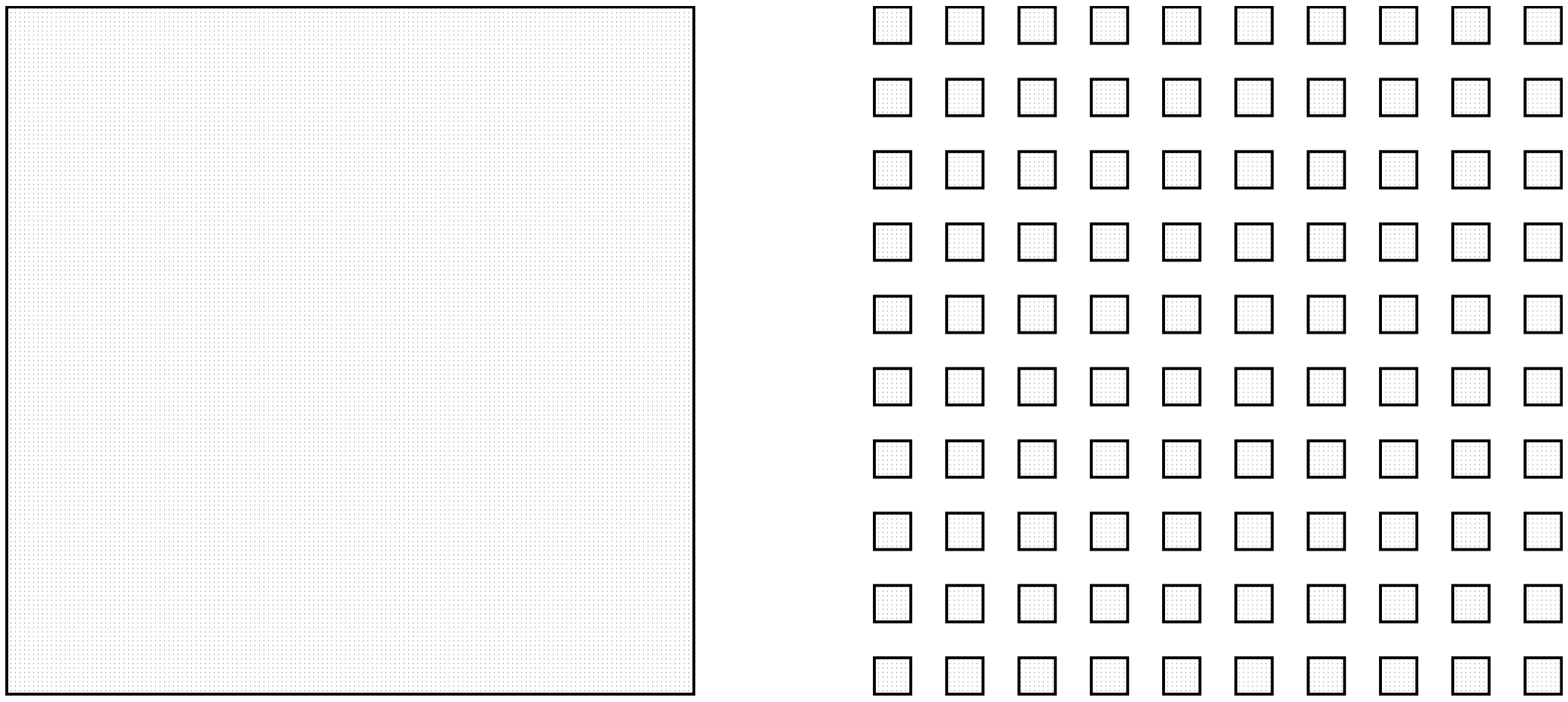

(a) Single unit @ reliability *R* (b) 100 units @ reliability *R*/10

Figure 10: With in a single unit (a), the aggregate reliability is the reliability of that one unit, but with 100 units (b) the aggregate is the same when the unit reliability is 10 times less. As a general rule if the unit reliability is *R* then the system reliability is $R \cdot N^{½}$. By allowing a significantly decreased unit reliability, while still maintaining high system reliability, significant cost savings in terms of O&M costs, avoided lost revenues, and customer outage costs can be achieved.

While this reliability estimate does not consider the implications of an interconnected system, it does suggest why having many small interacting machines leads to higher reliability than a few large units. The reliability impact of the power transmission and distribution network is much more difficult to estimate, but it is entirely reasonable to believe that the presence of rapidly responding loads can enhance reliability considerably. However, these considerations were not evaluated further in this report.

## 6.4 Customer Energy Efficiency Benefits

New GridWise technologies applied to the system (e.g., diagnostics at the system level to gain efficiency improvements) could result in additional energy efficiency and energy cost savings to customers. For example heating, ventilation and air-conditioning (HVAC) unit economizers fail between 20% and 50% of the time in modes that result in 20% to 30% unnecessary energy consumption. In this case, a savings of 20% to 30% to the customer is possible with smart appliance control.

Based on EIA data in the 2002 Annual Energy Outlook (EIA 2002), by 2020 the incremental improvement in energy efficiency enabled by GridWise technologies is expected to double the baseline of 2%, worth $2.6 billion to residential customers. For commercial customers, a 5% efficiency gain is expected to be worth $6.2 billion and to industrial customers a 10% efficiency gain would be worth $6.3 billion. Thus, customer energy efficiency benefits could reach $15.2 billion by 2020. These benefits would be worth an average of approximately $760 million per year and have a 20-year present value of $9 billion (see the Appendix, section A4).

## 6.5 Customer Outage Benefits

The benefits of reduced outage costs were addressed briefly in this assessment. Customer outage benefits would accrue from active grid management to reduce outage frequency and duration that would be enabled by GridWise technologies. We assumed a 50% reduction in outages where these technologies were implemented. In 2020, this reduction would be worth approximately $8.5 billion (Short et al. 2002). Over 20 years, this amounts to about $425 million incremental savings per year, having a PV of $5 billion.

# 7.0 Conclusions and Recommendation

Based on a preliminary cursory assessment, GridWise technologies are anticipated to offer remarkable benefit value if implemented as evolutionary changes in the Nation's electric power system. Evaluated conservatively, just a selected few of many possible benefits suggest that the GridWise vision could readily generate about $75 billion of benefit value over the next 20 years (see Table 1). When assessed less conservatively, Table 2 shows the total present value of these benefits essentially doubles.

Table 1: GridWise Benefits Based on Scenario 1

| Where Benefits Accrue | Scenario 1 Benefits ($B, 20-yr PV) | | | | | | |
|---|---|---|---|---|---|---|---|
| | Capacity Additions | Capacity Factor | Outages | End-Use Efficiency | Cost of Capital | Ancillary Services | Total |
| Transmission | 5.0 | | | | | 0.6 | 5.6 |
| Distribution | 22.0 | | 0.5 | | | | 22.5 |
| Generation | 19.0 | 2.5 | | | 11.0 | | 32.5 |
| Customer | | | 5.0 | 9.0 | | | 14.0 |
| Total | 46.0 | 2.5 | 5.5 | 9.0 | 11.0 | 0.6 | 74.6 |

Table 2: GridWise Benefits Based on Scenario 2

| Where Benefits Accrue | Scenario 2 Benefits ($B, 20-yr PV) | | | | | | |
|---|---|---|---|---|---|---|---|
| | Capacity Additions | Capacity Factor | Outages | End-Use Efficiency | Cost of Capital | Ancillary Services | Total |
| Transmission | 12.0 | | | | | 1.5 | 13.5 |
| Distribution | 56.0 | | 0.5 | | | | 56.5 |
| Generation | 49.0 | 6.4 | | | 11.0 | | 66.4 |
| Customer | | | 5.0 | 9.0 | | | 14.0 |
| Total | 117.0 | 6.4 | 5.5 | 9.0 | 11.0 | 1.5 | 150.4 |

Other benefits identified but not fully evaluated in this brief review have the potential to represent additional value in the $100 billion to $200 billion range. However, further study may find that some benefits such as the price volatility impact embed one or more of the benefits shown in Tables 1 and 2. The depth of the foregoing analysis did not resolve this issue. Therefore, to avoid the possibility of counting the same benefit more than once, we do not presently claim price volatility as an additional benefit. However, it represents an approach to "triangulating" the evaluation, and, as such, tends to confirm the order of magnitude of benefits accumulated by the alternative assessment techniques we have documented.

While implementation costs were not considered and the error band on the total benefit value is likely to be large (possibly ± $25 billion), the major conclusion of this exercise is that GridWise technologies have the potential for great economic value and should make a major contribution to transforming the present electric generation and delivery infrastructure into the power grid of the future.

A number of systemic benefits enabled by GridWise technologies were identified but not quantified. The transformed energy system will enable and reward services that improve the management of energy system. New restructured markets will pose fewer risks and have more participants, and thus provide a more dynamic and fine-tuned demand response, while creating greater confidence in the system's ability to respond to crises. New markets for new services will emerge, such as supply or demand aggregation, ancillary services based on load manipulation, and interdependency benefits. The new system will place emphasis on local solutions to problems such as capacity shortfall and grid congestion, and create incentives for just-in-time, right-sized, and right-sited infrastructure growth. Moreover, the incentives for overcapacity planning will disappear because of their disadvantageous economics. From environmental perspectives, markets for emissions trading and other environmentally desirable outcomes will be afforded a level playing field in these new markets. Most importantly, intermittent sources such as renewables will be accommodated more advantageously and in ways not possible in the past.

This study achieved its principal objective of establishing that potential benefits of the GridWise vision are, indeed, significant and worthy of further resolution. These results support our recommendation that the benefits and implementation costs of the concept should be subjected to more rigorous analysis.

# 8.0 References

ANSI. 1995. American National Standard for Electric Power Systems and Equipment-Voltage Ratings (60 Hertz), ANSI C 84.1-1995. Published by the National Electrical Manufacturers Association, Rosslyn, Virginia.

Braithwait, S. and A. Faruqui. 2001. Demand Response: The Ignored Solution to California's Energy Crisis. Public Utilities Fortnightly, March 15, 2001 Issue. Public Utilities Reports, Vienna, Virginia.

EIA. 2001. *Annual Energy Review 2000.* Report DOE/EIA-0384(2000) Energy Information Administration, U. S. Department of Energy, Washington, D.C.

EIA. 2002. *Annual Energy Outlook 2002.* Energy Information Administration, U. S. Department of Energy, Washington, D.C.

EIA. 2003(a). *Electric Power Annual 2001.* Energy Information Administration, U. S. Department of Energy, Washington, D.C.

EIA. 2003(b). *Annual Energy Outlook 2003.* Report DOE/EIA-0383(2003). Energy Information Administration, U. S. Department of Energy, Washington, D.C.

EPRI. 1993. Distributed Utility Valuation Project Monograph. Report TR-102807, Electric Power Research Institute, Palo Alto, California.

Hirst E. and B. Kirby. 1996. *Costs of Electric-Power Ancillary Services.* Electricity Journal, Volume 9, Issue 10, Pages 26-30. Elsevier Science B.V., Amsterdam, The Netherlands.

Hirst, E. and B. Kirby. 2001. Transmission Planning for a Restructuring U. S. Electricity Industry. Edison Electric Institute, Washington, D.C.

Lovins, A. B., E. K. Datta, T. Feiler, K. R. Rabago, J. N. Swisher, A. Lehmann and K. Wicker. 2002. *Small is Profitable.* Rocky Mountain Institute, Snowmass, Colorado.

NERC. 2002. *Reliability Assessment 2002 – 2011, The Reliability of Bulk Electric Systems in North America.* North American Electric Reliability Council, Princeton, New Jersey.

Short, T., B. Howe, W. Sunderman, A. Mansoor and P. Barker. 2002. *Analysis of Extremely Reliable Power Delivery Systems: A Proposal for Development and Application of Security, Quality, Reliability, and Availability (SQRA) Modeling for Optimizing Power System Configurations for the Digital Economy.* Product ID 051207. Electric Power Research Institute, Palo Alto, California.

Stoft, S. 2002. *Power Systems Economics: Designing Markets and Electricity.* Wiley Interscience, IEEE Press, Piscataway, New Jersey.

Willis, H. L. and W. G. Scott. 2000. *Distributed Power Generation: Planning and Evaluation.* Marcel Dekker, Inc., New York.

# APPENDIX A

Benefit Derivations

# Appendix A

The objective of this effort was to assess the hypothetical implementation of the GridWise concept to gain an order-of-magnitude perspective on the resulting benefits accruing nationally over a period of 20 years. This appendix presents additional details of benefit derivations that are not otherwise contained in the body of this report.

As a general approach, annual benefits were estimated on the basis of power grid statistics aggregated at the national level (see Table A1).

Table A1: Generation Scenario Assumptions and Sources

| ***Description*** | ***Value*** | ***Source*** |
|---|---|---|
| Total net generation | $3{,}792 \times 10^9$ kWh | EIA 2000, Figure 8.1 |
| US peak load | 685 GW (summer) | EIA 2000, Figure 8.14 |
| Net generation capacity | 819 GW<br>(includes utilities and NPPs) | EIA 2000, Figure 8.5 |
| Summer capacity margin | 14.6% | EIA 2000, Figure 8.14 |
| Average plant capacity factor | 53% | EIA 2003(a) Page 6 |
| Average cost of new generation | $600/kW | EIA 2003(b), Page 72 |

The present values (PV) of benefits were estimated using the following PV calculation:

$$PV = AB\frac{(1+I)^n - 1}{I(1+I)^n}(1+I)^{-t}$$

where:

$PV$ = present value
$AB$ = annual benefit
$I$ = annual discount rate (using 6%)
$n$ = number of years over which benefit accrues
$t$ = construction time of deferred plant (years).

For simplicity the term $(1 + I)^{-t}$ was made equal to unity. This, in effect, ignored the reduction in PV resulting from the time value of funds used during construction or AFUDC, which is generally allowed by regulators.

## A.1 Generation Benefits

Table A.1 lists data and sources used for estimating the national GridWise benefit resulting from enhanced utilization of generation assets. Two baseline scenarios of this type were considered. Each was based on the anticipation that the aggregate effect of implementing GridWise technologies would tend to flatten the load duration curve with the direct consequence of decreasing the rate at which new capacity needs to be added to meet future load growth and replace unit retirements.

**Scenario 1**: If the national load duration curve could be made essentially flat, then the total net generation recorded in 2000 would represent an average load of 432 GW ($3{,}792 \times 10^9$ kWh consumed over 8760 hours per year). With the 14.6% capacity margin existing in 2000, the available generation (including reserves) would need to be 496 GW [(432 x 1.146) GW] to provide the necessary protection. If the average plant capacity factor increased to 70%, 708 GW [(496/0.7) GW] of nameplate plant capacity would be necessary to maintain this capability. Comparing this to the 819 GW of net generation capacity in 2000 implies 111 GW of excess generation already exists if no load growth were to occur. Part of this excess capacity would be released by implementing the GridWise concept, thereby deferring some of the new construction needed for load growth and unit replacement. Assuming a 20-year implementation of GridWise technologies with a final penetration of 50%, the simple levelized benefit is the deferred value of half 111 GW over 20 years, or 2.8 GW per year. At $600/kW, this benefit is worth $1.7 billion ($2.8 \times 10^6$ kW x $600/kW) annually.

In this scenario, the assumption that average plant capacity factor reaches 70% reflects unit availability and utilization similar to that achieved by the Nation's coal-fired plants in 2001 (EIA 2003a). In 2001, average generation construction costs ranged from $536/kW for combined-cycle units to $1,367/kW for coal-fired steam plants (EIA 2003b). The $600/kW value used in Scenario 1 anticipates GridWise technologies would defer construction of mostly combined-cycle units. Assuming that GridWise technologies achieve only 50% penetration in 20 years is equivalent to the realistic expectation that a completely flat load duration curve would not be achievable in practice.

**Scenario 2**: Using the above basis more optimistically, if it is assumed that protecting a nearly flat load duration curve requires only a 5% reserve capacity margin, then the available generation capacity including reserves would need to be 454 GW [(432 x 1.05) GW] to provide 432 average MW. When combined with an improved capacity factor of 85%, this scenario would require only 534 GW [(454/0.85) GW of installed capacity. As above with 50% GridWise penetration, there would be 285 GW [819 – 534) MW] of excess capacity that could be deferred, which is equivalent to 7.1 GW per year worth $4.3 billion ($7.1 \times 10^6$ kW x $600/kW) annually for 20 years. An average capacity factor of 85% represents the majority of generating plants approaching the 89.4% capacity factor achieved by U. S. nuclear plants in 2001 (EIA 2003a).

The present values of GridWise enabled generation utilization enhancement estimated in the above manner are shown in Table A2.

Table A2: Present Value of GridWise Generation Capacity Benefits

| Baseline Scenario# | Present Value of Benefit ($ billion) |
|---|---|
| 1 - 50% market penetration<br>14.6% capacity reserves needed<br>70% plant capacity factor achieved | 19.4 |
| 2 - 50% market penetration<br>5% capacity reserves needed<br>85% plant capacity factor achieved | 49.3 |

## A.2 Transmission and Distribution Benefits

While T&D capacity costs have ranges dependent on the physical distance traversed, at an aggregate level the cost of T&D infrastructure can be expressed in terms of $/GW of load served. As discussed in Section 2.7, the cost of load growth was assumed to include $150 million/GW for transmission and $685 million/GW for distribution. It was further assumed that deferral of generation capacity by implementing the GridWise concept would similarly defer construction of the same amount of T&D capacity. The present value of deferred T&D construction was estimated in the same manner as the generation deferral benefit for each of the baseline scenarios described in Section A.1. These results are shown Tables A3 and A4.

Table A3: Present Value of GridWise Transmission Capacity Benefits

| Baseline Scenario # | Annual Benefit ($ billion) | Present Value ($ billion) |
|---|---|---|
| 1 | 0.42 | 5 |
| 2 | 1.07 | 12 |

Table A4: Present Value of GridWise Distribution Capacity Benefits

| Baseline Scenario # | Annual Benefit ($ billion) | Present Value ($ billion) |
|---|---|---|
| 1 | 1.92 | 22 |
| 2 | 4.86 | 56 |

## A.3 Ancillary Services Benefits

The grid-friendly load component of the GridWise concept can be expected to supply ancillary services that contribute to grid control and operation. Hirst and Kirby (1996) is a reference for average ancillary service costs in 1993, as summarized in Table A5.

Table A5: Cost of Ancillary Service in 1993

| ***Service*** | ***Cost (¢/kWh in 1993$)*** |
|---|---|
| Scheduling and dispatch | 0.018 |
| Generation reserves | |
| Load following | 0.038 |
| Reliability | 0.066 |
| Supplemental operating | 0.073 |
| Energy imbalance | 0.047 |
| Real-power losses | 0.122 |
| Voltage control | 0.051 |
| **Total Costs** | **0.414** |

Hirst and Kirby estimated the total cost of ancillary services in 1993 to be $12 billion per year. In 2000, the cost is estimated at 0.414 ¢/kWh times $3{,}792 \times 10^9$ kWh, or $15.7 billion per year without inflation. In the hypothetical limit, all GridWise connected loads can provide ancillary services to the extent that the GridWise concept makes them all dispatchable by the system operator. The benefit is estimated by taking the previous generation deferrals and considering the equivalent load provides 50% of the ancillary service's total value for all hours of the year.(4)

In Scenario 1, the first year benefit is 2.8 GW x 50% x 8,760 hours per year x 0.414 ¢ per kWh or $50.8 million. The present value of the benefit over 20 years is $0.6 billion.

In Scenario 2, the first year benefit is 7.1 GW x 50% x 8,760 hours per year x 0.414 ¢ per kWh or $128.8 million, with a 20-year present value of $1.5 billion.

## A.4 Energy Efficiency Estimates

The efficiency savings for customers were estimated from Annual Energy Outlook 2002 (EIA 2002) using projections for the years 2010 and 2020. These results are shown in Tables A6 and A7.

(4) Inflation of the Consumer Price Index has averaged about 2.5%/year since 1993. Because we found no information indicating that ancillary service costs have shown a similar inflation rate, the benefits were estimated in $1993 to be conservative.

Table A6: Energy Savings in 2010

| | Consumption (quad. Btu) | Price ($2000/MMBtu) | Cost of Electricity ($billion) | Energy Efficiency Improvement | Cost Savings ($billion) |
|---|---|---|---|---|---|
| Residential | 4.92 | 22.41 | 110 | 2% | 2.2 |
| Commercial | 5.03 | 19.87 | 100 | 5% | 5.0 |
| Industrial | 4.20 | 12.54 | 53 | 10% | 5.3 |
| Total | 14.15 | | 263 | | 12.5 |

Table A7: Energy Savings in 2020

| | Consumption (quad. Btu) | Price ($2000/MMBtu) | Cost of Electricity ($billion) | Energy Efficiency Improvement | Cost Savings ($billion) |
|---|---|---|---|---|---|
| Residential | 5.70 | 22.55 | 129 | 2% | 2.6 |
| Commercial | 6.13 | 20.33 | 125 | 5% | 6.2 |
| Industrial | 4.83 | 13.04 | 63 | 10% | 6.3 |
| Total | 16.66 | | 316 | | 15.1 |

The rate at which cost savings are projected to accrue average $0.76 billion/year over 20 years, although larger gains occur in the earlier years. Taking this rate as a lower bound, customer energy efficiency savings enabled by the GridWise concept have a conservative present value of $9 billion.

## A.5 Price Volatility Reduction Benefits

Another approach used to "triangulate" the impact of reduced price volatility achieved by demand response involved estimating the savings on a 50-GW system using 1990 supply data from Electric Reliability Council of Texas (ERCOT) and demand based on ELCAP models of consumer demand disaggregated into end uses (appliances and equipment). In this analysis we determined the price of power using a simple scarcity rent, i.e.,

$$offer = 30\left(1 - \frac{load}{capacity}\right)^{-2}$$

where *load*, *price*, and *capacity* are for a given hour. Note that *capacity* is given as the net available capacity, excluding reserve requirements.

We then estimated the demand response using the formula

$$response = 0.2 \cdot load\,\frac{offer}{200}$$

where 0.2 is the fraction of the load that responds (i.e., sheds) at the offer price of $200/MWh.

The quantity of power sold under these conditions is then given by

$$sold = load - response$$

and the price of the quantity sold is given by

$$price = 30\left(1 - \frac{sold}{capacity}\right)^{-2}.$$

The difference between the offer price (which corresponds to the price of power in an unresponsive demand scenario) and the price (for the responsive demand) is the savings per MWh. Thus the savings for a given hour is given by

$$savings = load(offer - price)$$

and the total savings for 1 year is

$$total = \sum_{8760} savings.$$

In the scenario studied, the savings were $694 million, and the peak load was reduced from 42 GW to 36 GW. Extrapolating this result to 1,000 GW installed capacity results in $14 billion in annual savings with a present value of approximately $160 billion.

# Distribution

## Offsite

A. Ipakchi
ALSTOM
101 Metro Drive, Suite 750
San Jose, CA 95110

R. Guild c/o Ali Ipakchi
ALSTOM
101 Metro Drive, Suite 750
San Jose, CA 95110

J. Berst
The Athena Institute
15127 NE 24th Suite 358
Redmond, WA 98052

M. Hoffman
Bonneville Power Administration
P.O. Box 3621
Portland, OR 97208-3621

T. Oliver
Bonneville Power Administration - PNG
P.O. Box 3621
Portland, OR 97208-3621

R. Hoffman
Consultant to the California Energy Commission
847 Mountain Blvd.
Oakland, CA 94611-0375

T. Surles
California Energy Commission
Public Interest Energy Research (PIER)
1516 Ninth Street, MS-29
Sacramento, CA 95814

P. Wang
Concurrent Technologies Corp.
320 Pitt Way
Pittsburgh, PA 15238

D. Samuel
IBM
General Mgr, Global Energy & Utilization Ind.
404 Wyman Street
Waltham, MA 02454

R. DeBlasio
Program Manager
National Renewable Energy Laboratory
1617 Cole Blvd.
Golden, CO 80401-3393

G. Nakarado
National Renewable Energy Laboratory
1617 Cole Boulevard
Golden, CO 80401-3393

M. Dworkin
President
New England Conference of
Public Utilities Commissioners, Inc.
One Eagle Square, Suite 514
Concord, NH 03301

P. Harris
President
PJM Interconnect
Valley Forge Corporate Center
955 Jefferson Ave.
Norristown, Pennsylvania 19403-2497

W. Baer
Science and Technology Division
RAND
1700 Main Street, P.O. Box 2318
Santa Monica, CA 90407-2138

B. Fulton
Doctoral Fellow
The RAND Graduate School
1700 Main Street, P.O. Box 2318
Santa Monica, CA 90407-2138

S. Cherian
Spirae, Inc.
4405 Gray Fox Rd.
Fort Collins, CO 80526

S. Hauser
DigitalCity™ Business Unit
UAI
307 Wynn Drive NW
Huntsville, AL 35805

S. Gehl
Strategic Technology
EPRI
3412 Hillview Avenue
Palo Alto, CA 94304-1395

R. Ambrosio
IBM T.J. Watson Research Center
P.O. Box 218
Yorktown Heights, NY 10598-0218

J. Glotfelty TD-1
Office of Electric Transmission and Distribution
U.S. Department of Energy
1000 Independence Avenue S.W.
Washington, DC 20585

E. Lightner EE-2D
Office of the Distributed Energy and Electricity Reliability
U.S. Department of Energy
1000 Independence Avenue S.W.
Washington, DC 20585

W. Parks EE-2D
Office of the Distributed Energy and Electricity Reliability
U.S. Department of Energy
1000 Independence Avenue S.W.
Washington, DC 20585

# Onsite

**DOE Richland Operations Office**
K. Williams K8-50

**Pacific Northwest National Laboratory**

| | |
|---|---|
| M. J. Lawrence | K7-73 |
| L.D. Kannberg | K5-02 |
| S.A. Sande | K5-02 |
| M.C. Kintner-Meyer | K5-16 |
| R.G. Pratt | K5-16 |
| L.A. Schienbein | K5-20 |
| W.M. Warwick | BPO |
| C.H. Imhoff | K5-02 |
| J.G. DeSteese (2) | K5-20 |
| AM. Borbely-Bartis | K6-05 |
| Hanford Technical Library | P8-55 |